# Instant Response of Live HeLa Cells to Static Magnetic Field and Its Magnetic Adaptation

Sufi O Raja and Anjan K Dasgupta[*]

Department of Biochemistry, University of Calcutta, 35 Ballygunge Circular Road, Kolkata-700019, India

[*]E-mail: adbioc@caluniv.ac.in

## Abstract

We report Static Magnetic Field (SMF) induced altered sub-cellular streaming, which retains even after withdrawal of the field. The observation is statistically validated by differential fluorescence recovery after photo-bleaching (FRAP) studies in presence and absence of SMF, recovery rate being higher in presence of SMF. This instant magneto-sensing by live cells can be explained by inherent diamagnetic susceptibility of cells and alternatively by spin recombination, e.g., by the radical pair mechanism. These arguments are however insufficient to explain the retention of the SMF effect even after field withdrawal. Typically, a relaxation time scale at least of the order of minutes is observed. This long duration of the SMF effect can be explained postulating a field induced coherence that is followed by decoherence after the field withdrawal. A related observation is the emergence of enhanced magnetic susceptibility of cells after magnetic pre-incubation. This implies onset of a new spin equilibrium state as a result of prolonged SMF incubation. Lastly, translation of such altered spin states to a cellular signal that leads to an altered sub-cellular streaming, probable intracellular machineries for this translation being discussed in the text.

# 1. Introduction

There are several reports on effect of SMF on cellular and genetic level [1, 2]. While the exact triggering stimulus is unclear, the SMF seems to initiate a number of events that modulates the cell shape, motility and energetic [3, 4]. Simplistic thermodynamic arguments can hardly explain the observed effect, particularly its bulk level manifestation. For example an one Tesla field (~$10^{-7}$ ev) is several order less in magnitude than thermal fluctuations (0.04 ev) operative at normal temperature. Many of the observations reported earlier, and reported by us in this paper, use SMF of even lower values. SMF can resolve the degeneracy of electronic level resulting in an enhanced fluorescence quantum yield. At low temperature such SMF induced enhancement of fluorescence is explained by the Radical Pair mechanism that involves the Zeeman interaction and hyperfine coupling [5]. The SMF induced enhanced fluorescence is also employed in imaging platform to enhance the image contrast of fixed cells [6]. But, the effect is observed only when the field is present.

On the other hand, geomagnetic field sensing by migratory birds evokes quantum coherence at wet conditions and that too at physiological temperature [7]. Long term coherence between different spin populations within a specialized protein, known as cryptochrome, is the basic reason of geomagnetic field sensing the sensing being operative even in presence of noise or thermal fluctuations [8, 9]. Our concern is the question of global validity of this mechanism of magnetic field sensing in noisy systems. The question is do we always need a cryptochrome like protein to explain such static magnetic field effects or, is there any related effect that is not explainable in terms of this spin recombination mechanism. Non-classicality of quantum phenomenon and possible close coupling between thermal energy and excitation behavior has recently been discussed in a recent paper [10]. The key question according this author is the validity of quantum behavior at conditions where the conventional coherence behavior is unlikely to exist and thermal fluctuations play a dominating role.

We have shown the live cells responds to a moderate magnitude of SMF, and the signal thus obtained is translated into an altered sub-cellular flow. This translation being a slower process, the memory of the SMF exposure is retained at time scale not explainable by any conventional quantum mechanical formalism.

The alternative explanation of magneto-sensing by diamagnetic anisotropy has two major obstacles. There are reports on diamagnetic levitation [11] using very high magnetic field set up [12, 13]. In other words such effect at relatively moderate SMF strength (say at 0.5T) is unlikely to occur. Secondly, this argument is insufficient to explain the long term relaxation behavior as we have observed.

Again, the sources of diamagnetic anisotropy in biological systems are amino acids, lipids, peptide bonds [14]and their ordered structures like secondary structures of proteins [15] and organelles (thylakoid) [16]. A single diamagnetic molecule may not overcome the thermal energy to show any macroscopic effect in response to SMF. But, ordered biological structures create giant diamagnetic anisotropy, which can beat the thermal energy [15]. Proteins like tubulin [17] and fibrin [18] can be oriented using SMF. The instant response of live cells to SMF may be due to occurrence of diamagnetic torque as cells are in dynamic equilibrium. While the diamagnetism alone cannot explain the memory like effect, the possibility of such components translating the quantum signal to the bulk cellular milieu cannot be ruled out. The paper shows some simple experimental proof of this quantum signal translation machinery in the cell.

**2. Materials and Methods:**

*2.1. Cell line maintenance and transfection:*

HeLa cells were procured from NCCS, India. Cells are cultured in DMEM, high glucose (Gibco 12800-058) with 10% heat inactivated FBS and the media is supplemented with 1X antibiotic (PenStrep Gibco). Cells were maintained at 5% $CO_2$ (Carbon di Oxide) environment. For subsequent passages, cells were dislodged using 0.25% Trypsin EDTA (1X) (GIBCO 25200). HeLa cells were transfected by pEGFP-N3 plasmids using calcium phosphate transfection protocol as described previously [19].

*2.2. Confocal imaging experiment:*

The plates were observed directly under inverted Confocal microscope (Olympus, FV1000) using 60X objective lens. Experimental set up has been shown in Figure 1. 50 mTesla magnetic field was created using a Neodymium (Nd) magnet (0.5 Tesla), purchased from Rare Earth Magnetics (India). We placed the magnet without disturbing the plate at various time intervals. Field strength was measured by a standard Gauss-meter.

*2.3. Time lapse imaging experiments:*

Time lapse imaging was performed in inverted confocal microscope (Olympus, FV1000). Six time lapse images were acquired at 1 min time interval. Image 1 and 2 were acquired in absence of magnetic field, image 3 and 4 were acquired in presence of field and last two images (image 5 and 6) were acquired after withdrawal of the field. We performed control experiment in the same order using an aluminium foil.

*2.4. Fluorescence Recovery After Photobleaching (FRAP) studies:*

FRAP experiments were performed in absence and in presence of magnet using an appropriate macro. We selected two regions. Using 80% laser power we bleached one region, keeping other region as reference. Then we captured 11 time lapse images at 5 sec time interval in absence and in presence of magnet.

*2.5. Mitochondria staining and imaging:*

Mitotracker Red CMXRos came with Image-iT™ LIVE Mitochondrial and Nuclear Labeling Kit (I34154), molecular probes, Invitrogen. Staining solution was prepared according to the supplied protocol. Final working concentration was 25 nM.

*2.6. Image analysis:*

Gray scale and pseudo-coloured difference images were constructed using Matlab R14 (MATHWORKS, USA). 16 bit image processing and intensity histogram analysis were also performed. The fluorescence recovery kinetics was calculated from time lapse images. The detailed m-scripts are provided in the supplementary section.

**3. Results and Discussions:**

We performed time lapse and FRAP experiments in absence and presence of SMF using the experimental set up shown in Figure 1. We captured 6 time lapse images at intervals of miniute. Image 1 and 2 are in absence of SMF, image 3 and 4 are in presence of SMF and image 5 and 6 are after withdrawal of the field. Figure 2a shows the original image of GFP transfected HeLa cells. Figure 2b is the difference image (image2-image1) without field, Figure 2c is the difference image (image4-image1) with field and Figure 1d is the difference image (image6-image1) after withdrawal of the field. Figure 2c and 2d are different from Figure 2b. The SMF induced altered sub-cellular pattern are clearly visible from the difference images. In gray scale difference images, the white pixel indicates enhancement of fluorescence intensity or protein inflow in the plane of observation at that region and the black pixel means reduction of fluorescence intensity (bleaching) and/or protein outflow from that region.

In control sets, where magnetic field was absent, some pattern unlike Figure 1 was observed, but the rate of photo-bleaching is also high (see Figure 3). The observed pattern may be due to differential bleaching with time. In this context we can conclude that SMF reduces the rate of bleaching also (see Figure 2 and 3). See supplementary section (Figure S1 and S2) for another set of time lapse imaging in absence (control) and presence of SMF using same experimental set up.

As in difference image, the white pixels are found in some specific areas within cells we constructed a pseudo colored image to observe the region within cells susceptible to SMF. In the pseudo colored image the Red plane is assigned for enhancement of pixel density, the Green plane is for the original image (image1) and Blue plane is assigned for reduction of pixel density. Figure 4 shows the pseudo colored image, where yellow (red + green) pixels indicate the enhancement of fluorescence and cyan (green + blue) pixels indicate reduction of fluorescence intensity. Figure 4a shows the original 16 bit Confocal image of GFP transfected HeLa cell. Before introduction of magnet (Figure 4b) the yellow pixel distribution is random. But after placing the magnet the yellow pixels distribution become clustered (see Figure 4c) and retains after withdrawal of the field (see Figure 4d).

For further validations of SMF induced altered sub-cellular streaming we performed FRAP experiments in absence and presence of SMF. Figure 5a shows that the fluorescence recovery kinetics in presence of magnet is higher than in absence of magnetic field. As FRAP studies used for sub-cellular mobility assay [20], our results clearly indicate that the SMF recruits higher protein (higher mobility) in the plane of observations. Figure 5b shows the box plot of six different sets of FRAP experiments in absence and presence of field, where the mean values of the boxes also indicates the enhanced fluorescence recovery rate in presence of field.

As mitochondria are the energy hub of cells and any cellular movement needs energy, we stained mitochondria of live HeLa cells to observe whether there is any response to SMF in image platform. We performed time lapse imaging in absence and presence of SMF in same set up. We found significant decrease in fluorescence intensity of Mitotracker Red upon introduction of SMF. In this context we should mention that reduction of fluorescence intensity in presence of SMF cannot be explained by RP mechanism (enhancement was expected [6]). We observed increase in fluorescence intensity after withdrawal of the SMF. Figure 6 shows the 16 bit Confocal images of Mitotraker Red stained mitochondria of live HeLa cells in absence (a), presence (b) and after withdrawal (c) of the field. Figure 6d shows the overall intensity of 16 bit images in absence, presence and after withdrawal of the field. The intensity distribution in absence and after withdrawal of the field is similar. But, in presence of field we observed a cluster of low pixel density with high frequency (see Figure 6d). Now Mitotracker Red stains only polarized mitochondria and decrease of intensity indicate depolarization of mitochondria [21]. From Figure 5 we can say that mitochondria get depolarized upon exposure of SMF. Recently we have shown the cell specific differential effect on DNA fragmentation and ROS generation upon long term SMF incubation (accepted for publication in Cancer Nanotechnology). Now mitochondrial membrane potential and intracellular tubulin dynamics are closely coupled. As the number of free tubulin increases, the mitochondrial membrane potential decreases (depolarization) due to binding of free tubulin with Voltage Dependent Anion Channels (VDAC) of mitochondria [22]. Hence, the observed SMF induced altered sub-cellular pattern is due to diamagnetic torque induced altered tubulin polymerization/de-polymerization dynamics.

The most interesting part is the retention of the altered streaming after withdrawal of the field. Due to relaxation time scale issue the memory effect cannot be explained by RP mechanism or inherent diamagnetic anisotropy of the cells. RP mechanism can only operate when the magnetic field is present. So we can rule out the first possibility. Now, according to classical diamagnetic relaxation theory the magnetization should be vanished just after withdrawal of the field (instantaneous). But, chaotic systems and plasma like systems can show a very large value of relaxation time, sometimes near infinity (altered equilibrium state) [23, 24]. Diamagnetism has also an electronic spin based quantum mechanical explanation. A giant diamagnetic anisotropy within biological structures originates due to spatial coherence between different spin states. Macroscopic quantum coherence within protein due to quantum dipole oscillation was first proposed by Frohlich [25, 26]. The role of tubulin in consciousness through quantum super-position was shown by Penrose and Hameroff [27]. Live cells can be considered as plasma state due to high degree of ordering of water molecules on cytoskeleton actin, microtuble and other structures [28]. This type of structural orderliness is resulted from coherent spin states and their dynamic nature makes the cell a coherent system. Now external spin perturbation may introduce decoherence in the system. If the coherency is already lost then cells cannot sense the magnetic field anymore. We did not observe SMF induced sub-cellular patterning in paraformaldehyde fixed cells (see Figure S3). Live cells will try to overcome the spin perturbation through triggering of different signaling cascade. We found that the instant response is altered sub-cellular streaming and depolarization of mitochondrial membrane. The membrane potential restored after withdrawal of the field, but altered sub-cellular streaming retains. Change in mitochondrial membrane potential can trigger different downstream signaling cascade, which is cell specific and results in the reported effects on cellular level upon long term magnetic incubation. We observed that long term magnetic incubation makes the cells more sensitive to SMF (memory of magnetic exposure). Figure 7e shows that when field is present the overall shape changes and after withdrawal initial shape restores (see Figure 7f). But, in control set (absence of field) time dependent bleaching is observed (see the arrow in the upper panel of Figure 7). That means prolonged magnetic incubation altered the inherent coherent state to a different state, which have a magnetic memory. Hameroff also proposed that if coherency is lost, a normal cell may become cancerous (alternative coherent state) [29]. We showed that external spin perturbation can also introduce alternative spin states with a large decoherence time scale and in prolonged exposure of the field the alternative states become stable (more susceptible to SMF [see Figure 7]). This alternative state is different from the cell specific original state and is the reason of cell specific effects of long term SMF incubation.

## 4. Conclusions

We have shown that the magneto-sensing is not limited to systems containing Cryptochrpme like specialized protein. A live cell can also sense magnetic field in terms of altered sub-cellular streaming. The differential streaming pattern is validated higher rate of fluorescence recovery in presence of SMF. The retention of altered sub-cellular streaming after withdrawal of field can be explained by SMF induced alternative coherent spin state, follows by a large decoherence time (order of the minutes). This long relaxation time scale or magnetic memory cannot be explained by existing theory like RP mechanism or diamagnetic susceptibility. The SMF induced alternative coherent state can be stable upon long term SMF incubation, as then cells become more susceptible to SMF. The observed results also indicate the cross talk between altered sub-cellular dynamics and mitochondria, as SMF resulted in depolarization of mitochondrial membrane potential. The exact triggering of downstream signaling cascade is still needs further research. But, we can say that the instant response to SMF is the altered sub-cellular dynamics, which is coupled to cellular energetics and can result in cell specific differential late effects (reported) through triggering of cell specific different signaling cascades.

## Acknowledgement

Part of this work was supported by ICMR, India (No: 35/24/2010/BMS-NANO). We like to acknowledge DBT-IPLS instrument facilities in CU and Ms Boni Halder for technical assistance in confocal imaging.

## Figures

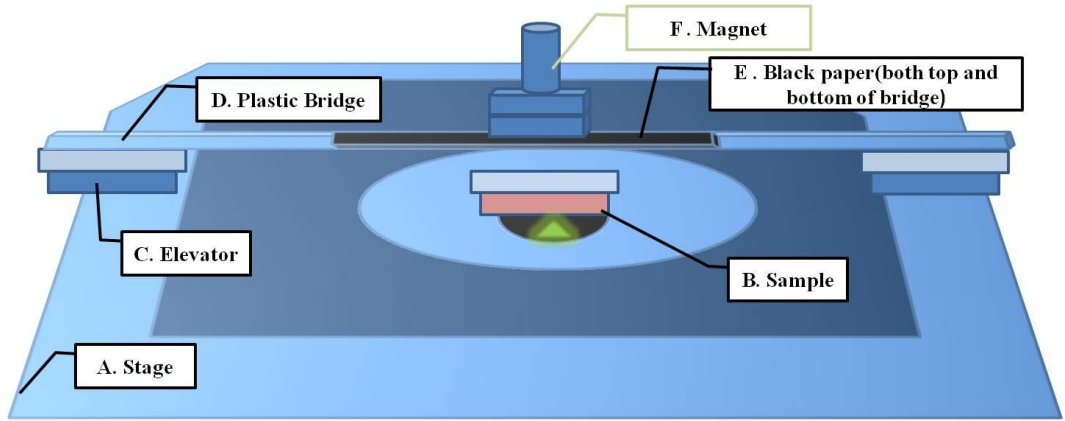

**Figure 1:** Schematic representations of the imaging set up using a magnet. The mounted plastic bridge was used so that the stage does not get disturbed during placing and withdrawal of the magnet.

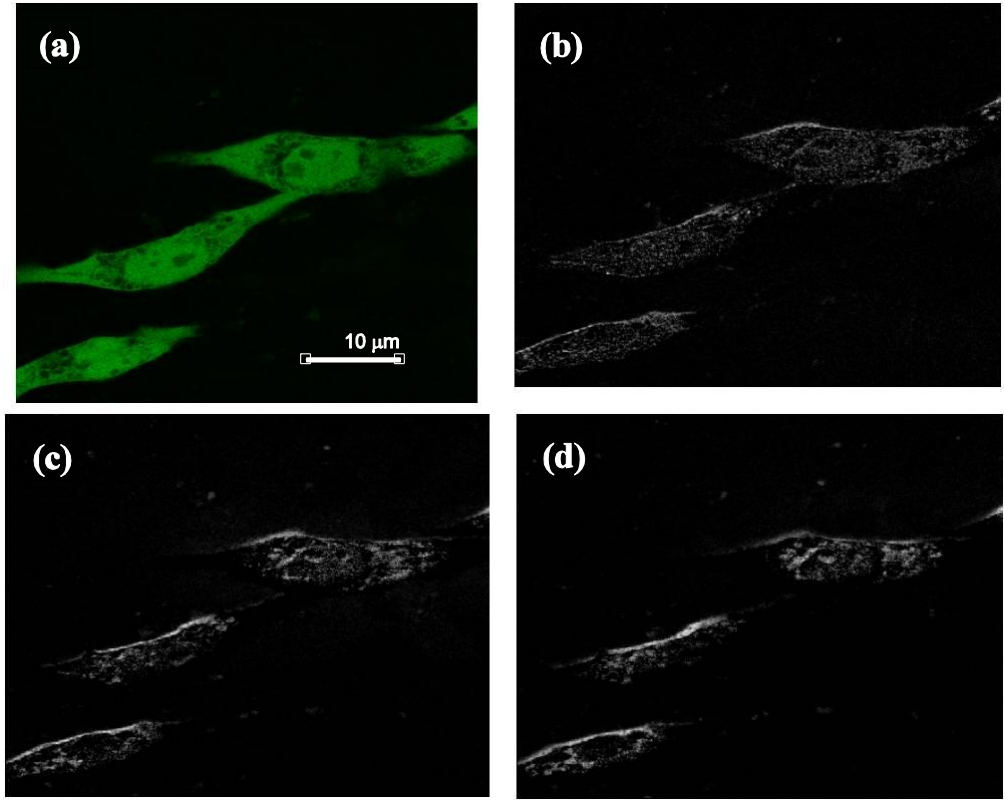

**Figure 2:** (a) 16 bit confocal image of GFP transfected live HeLa cells captured at 60X magnification with 2X optical zoom. Six time lapse images were captured at 1 min time interval. Image 1 and 2 are in absence of magnet, image 3 and 4 are in presence of magnet and image 5 and 6 are after withdrawal of the magnet. Grayscale difference image (b): image 2-image 1, (c): image 4-image 1and (d): image 6-image1.

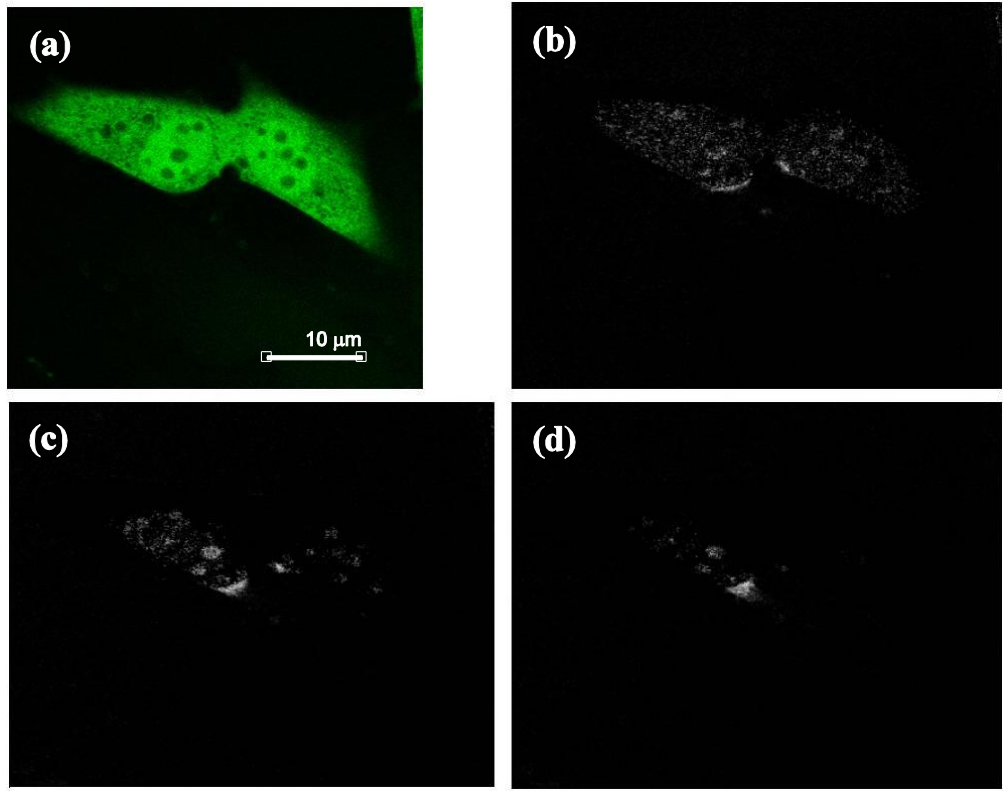

**Figure 3:** (a) 16 bit confocal image of GFP transfected live HeLa cells captured at 60X magnification with 2X optical zoom. Six time lapse images were captured at 1 min time interval. All six images were captured in absence of SMF. Grayscale difference image (b): image 2-image 1, (c): image 4-image 1 and (d): image 6-image1.

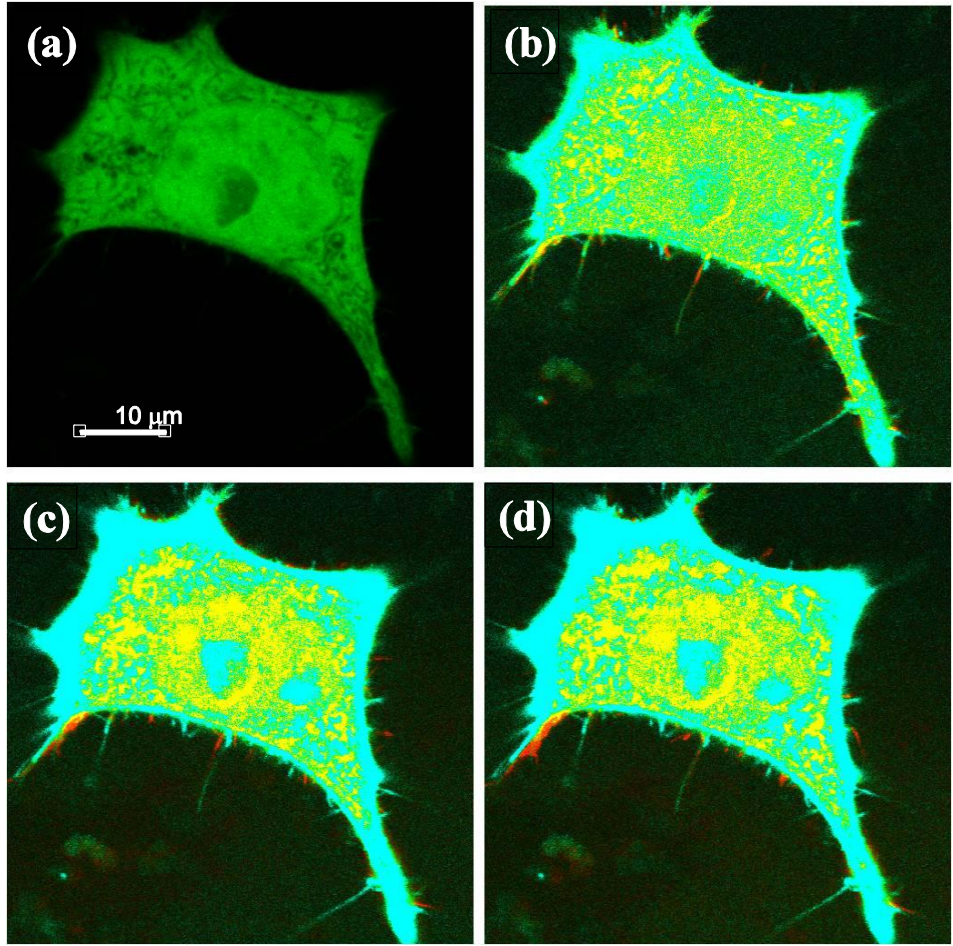

**Figure 4:** (a) 16 bit confocal image of GFP transfected live single HeLa cell at 60X magnification with 4X optical zoom. Six time lapse images were taken with 1 min interval. Image 1 and 2 are in absence of field, image 3 and 4 are in presence of field and image 5 and 6 are after withdrawal of the field. Pseudo colored difference images: image 2-image 1 (b), image 4-image 1 (c) and image 6-image1. In pseudo colored images 'R' plane is assigned for pixel enhancement, 'G' plane is for original image and 'B' plane is for pixel reduction. Here yellow (R+G) means enhancement and cyan (G+B) means reduction.

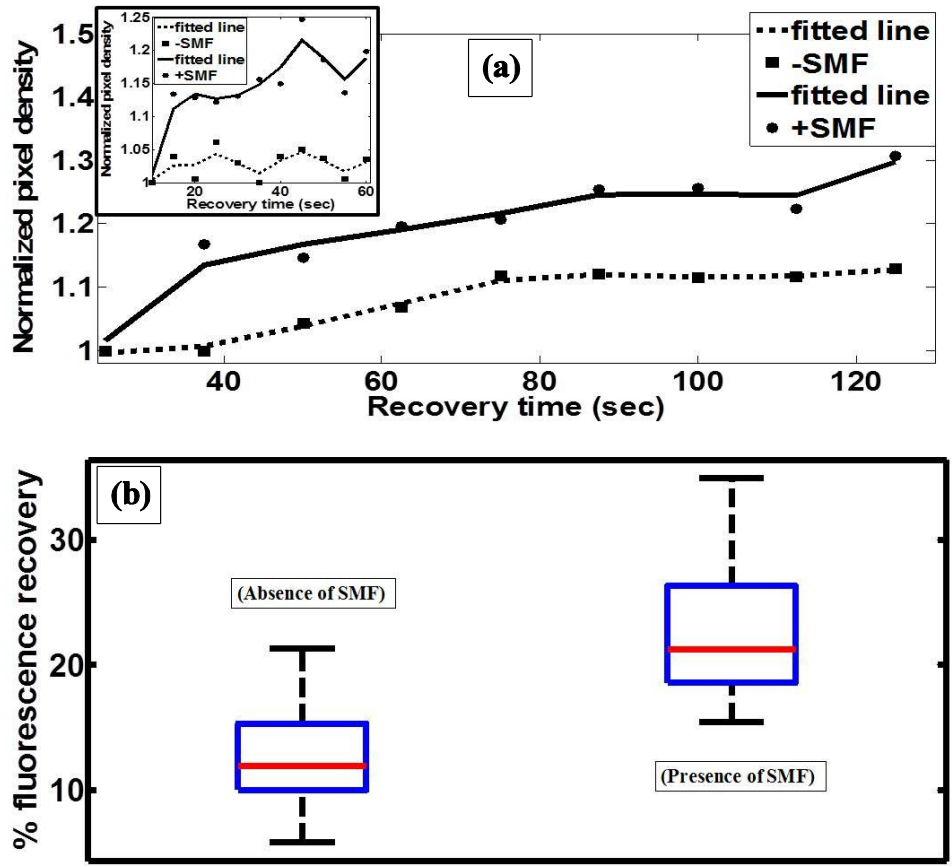

**Figure 5:** (a) fluorescence recovery kinetics for 120 mins after bleaching in absence (filled square) and in presence (filled circle) of magnetic field. Inset shows the same for 60 mins recovery kinetics. The solid and broken lines are spline fitted lines. (b) Box plot for six different sets of recovery kinetics in absence and presence of magnetic field.

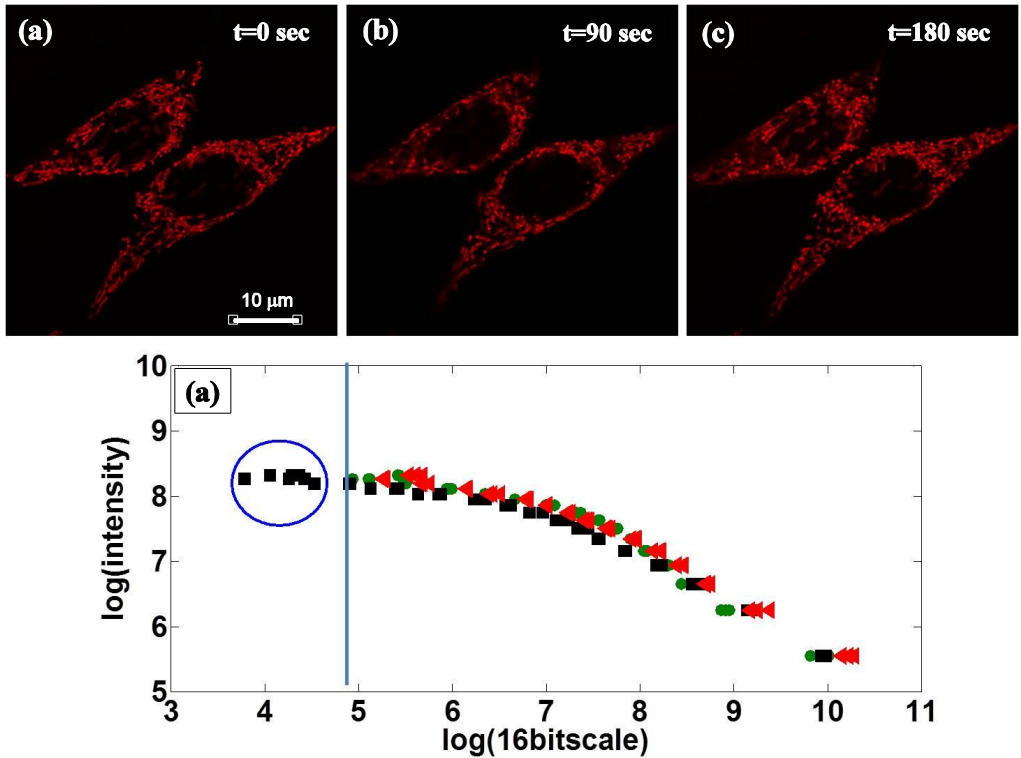

**Figure 6:** Mitotracker Red stained mitochondria of live HeLa cells (a) in absence, (b) in presence and (c) after withdrawal of the magnet. (d) Intensity distribution of images of a, b and c. X axis is the 16 bit pixel density and Y axis is the frequency of the pixels. Nine time lapse images were acquired using 30 sec of time interval. First three images were in absence (red points), next three images were in presence (black points) and last three images were acquired after withdrawal of the magnetic field (green points).

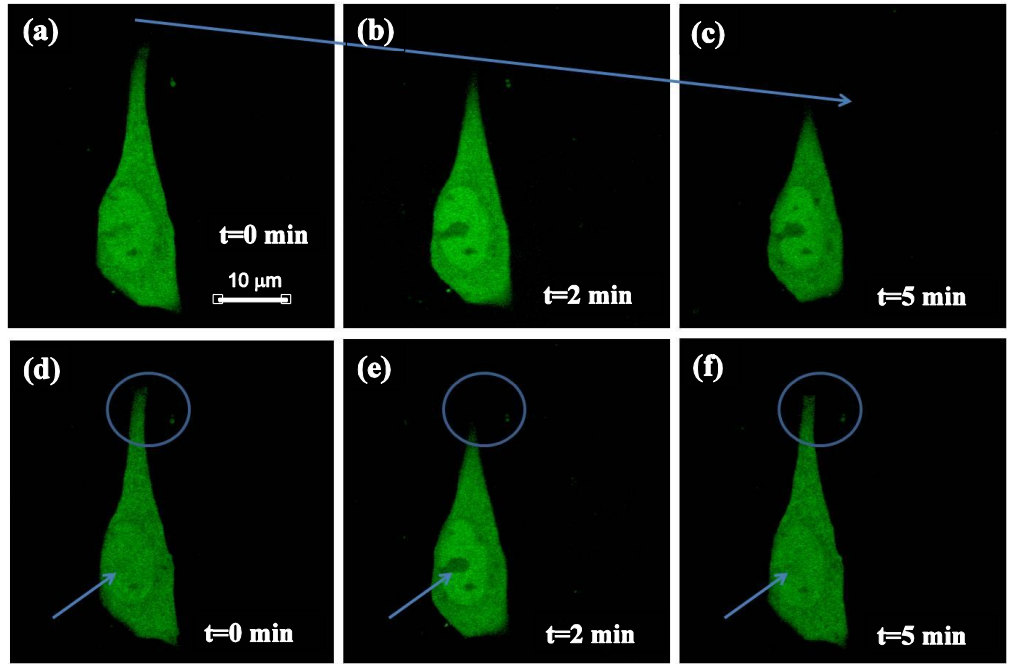

**Figure 7:** 16 bit Confocal images of live HeLa cells, captured after 24 hrs of magnetic (70 mT) incubation. Upper panel (a, b and c) is for control set, where no magnetic field is present. Lower panel shows the images of same cell in absence (d), in presence (e) and after withdrawal of the field (f). The arrows and circles show the changed regions.

# Supplementary Informations

# Instant Response of Live HeLa Cells to Static Magnetic Field and It's Magnetic Adaptation

Sufi O Raja and Anjan K Dasgupta[*]

**Fluorescence Recovery After Photobleaching (FRAP) analysis using Matlab, R2009b (MATHWORKS, USA):**

We analyzed the temporal evolution of fluorescence intensity after bleaching of a fixed Region of Interest (ROI) in absence and in presence of magnet. The script is as follows;

```matlab
function [tt,frap1,frap2,ffb1,ffb2]=raja_frap_sep(pth,pth1,fmt);
[fname,cw]=imxtract(pth,fmt);
;
[fname1,cw1]=imxtract(pth1,fmt);

fmt='tif'

l=length(fname);
% Laser irradiation regimes

box1=[304 280 10 10];% Absence

box2=[240 253 12 13];% Presence

% crop the images
% find out total pixel /unit area
% following are reference bboxes where no bleaching occured
boxref1=[161 200 20 18]; %absence
boxref2=[393 361 13 14];%presence

frap1=[];
frap2=[];
ffb1=[];
ffb2=[];
tt=[];

for i=1:l
    tt=[tt;i]
    s1=imcrop(cw{i},box1);
    fb1=imcrop(cw{i},boxref1);

    s2=imcrop(cw1{i},box2);
```

```
    fb2=imcrop(cw1{i},boxref2);

    s1=double(s1);
    s1=mean(mean(s1));

    s2=double(s2);
    s2=mean(mean(s2));

    fb1=double(fb1);
    fb1=mean(mean(fb1));

    fb2=double(fb2);
    fb2=mean(mean(fb2));

        frap1=[frap1 s1];
        ffb1=[ffb1 ;fb1];

        frap2=[frap2 s2];
        ffb2=[ffb2 ;fb2];

end
```

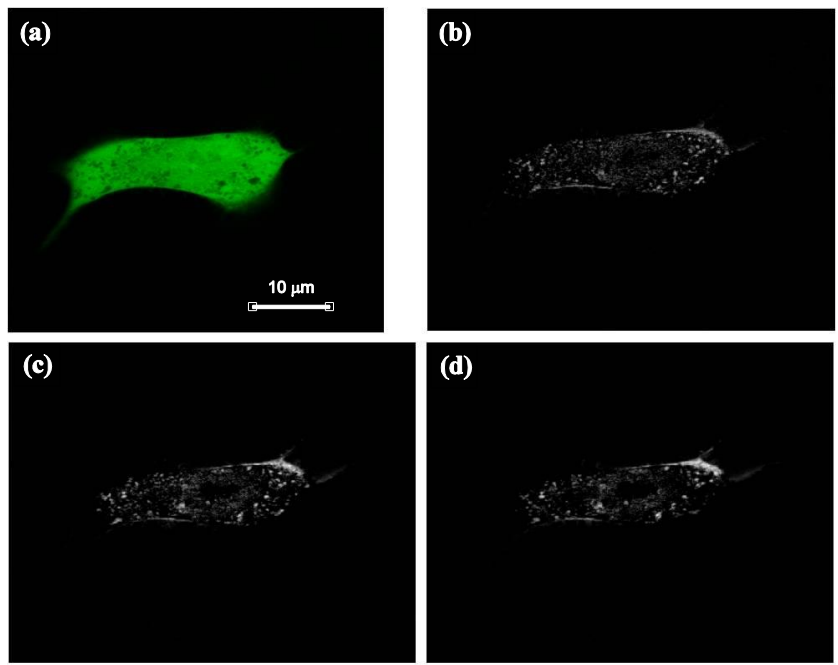

Figure S1: (a) 16 bit confocal image of GFP transfected live HeLa cells captured at 60X magnification with 2X optical zoom. Six time lapse images were captured at 1 min time interval. Image 1 and 2 are in absence of magnet, image 3 and 4 are in presence of magnet and image 5 and 6 are after withdrawal of the magnet. Grayscale difference image (b): image 2-image 1, (c): image 4-image 1and (d): image 6-image1.

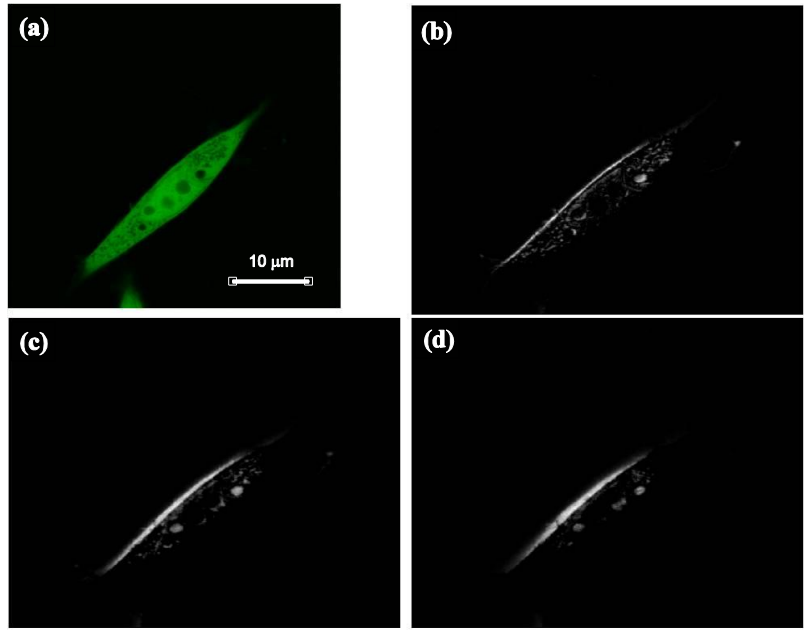

Figure S2: (a) 16 bit confocal image of GFP transfected live HeLa cells captured at 60X magnification with 2X optical zoom. Six time lapse images were captured at 1 min time interval. All six images were captured in absence of SMF. Grayscale difference image (b): image 2-image 1, (c): image 4-image 1and (d): image 6-image1.

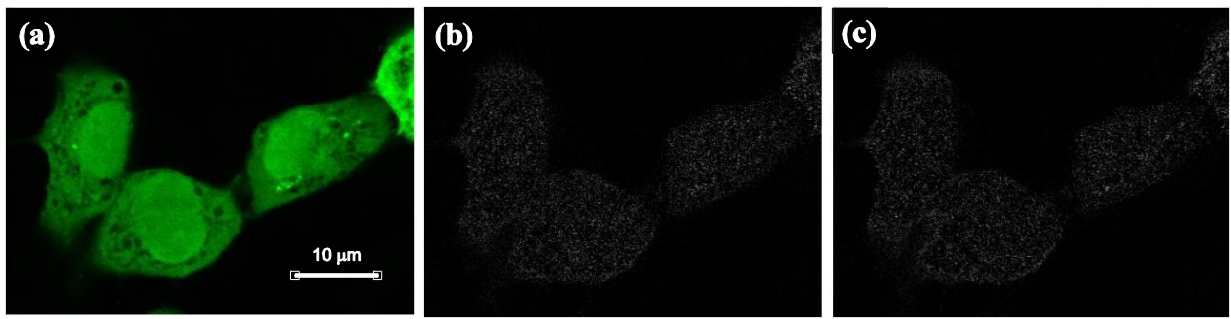

Figure S3: (a) 16 bit confocal image of GFP transfected paraformaldehyde fixed HeLa cells captured at 60X magnification with 2X optical zoom. Six time lapse images were captured at 1 min time interval. Image 1 and 2 are in absence of magnet, image 3 and 4 are in presence of magnet and image 5 and 6 are after withdrawal of the magnet. Grayscale difference image (b): image 2-image 1, (c): image 4-image 1.